\documentclass[%
 reprint,
superscriptaddress,
 amsmath,amssymb,
 aps,
prb,
]{revtex4-2}

\usepackage{graphicx}
\usepackage{overpic}   
\usepackage{multirow}
\usepackage{dcolumn}
\usepackage{bm}
\usepackage{color}
\usepackage[colorlinks=true, linkcolor=black, citecolor=black, urlcolor=black]{hyperref}
\usepackage{tikz}
\usepackage{upgreek}
\usepackage{float}
\usepackage{xcolor,soul}
\sethlcolor{yellow!35} 
\soulregister\cite7  
\usepackage[most]{tcolorbox}
\newtcolorbox{revhl}{
  enhanced, breakable,
  frame hidden,
  colback=yellow!35,
  boxrule=0pt, colframe=yellow!35,
  boxsep=0pt, left=0pt, right=0pt, top=0pt, bottom=0pt,
  arc=0pt, outer arc=0pt,
  enlarge left by=0pt, enlarge right by=0pt,
  nobeforeafter, 
  before upper={%
    \par
    \setlength{\parindent}{5pt}
    \parshape=2
      \parindent \dimexpr\linewidth-\parindent\relax
      0pt        \linewidth
    \leavevmode\indent
  },
  after upper={\par\parshape=0}
}

\begin{document}

\preprint{APS/123-QED}

\title{Quantum Transport in Ultrahigh-Conductivity Carbon Nanotube Fibers}

\author{Shengjie Yu}
\thanks{These authors contributed equally to this work.}
\affiliation{Department of Electrical and Computer Engineering, Rice University, Houston, Texas 77005, USA}
\affiliation{Applied Physics Graduate Program, Smalley-Curl Institute, Rice University, Houston, Texas 77005, USA}
\affiliation{Carbon Hub, Rice University, Houston, Texas 77005, USA}

\author{Natsumi Komatsu}
\thanks{These authors contributed equally to this work.}
\affiliation{Department of Electrical and Computer Engineering, Rice University, Houston, Texas 77005, USA}
\affiliation{Carbon Hub, Rice University, Houston, Texas 77005, USA}
\affiliation{Department of Bioengineering, University of Illinois Urbana–Champaign, Urbana, Illinois 61801, USA}

\author{Liyang Chen}
\affiliation{Applied Physics Graduate Program, Smalley-Curl Institute, Rice University, Houston, Texas 77005, USA}
\affiliation{Department of Physics and Astronomy, Rice University, Houston, Texas 77005, USA}

\author{Joe F. Khoury}
\affiliation{Department of Chemical and Biomolecular Engineering, Rice University, Houston, Texas 77005, USA}

\author{Nicolas Marquez Peraca}
\affiliation{Department of Physics and Astronomy, Rice University, Houston, Texas 77005, USA}

\author{Xinwei~Li}
\affiliation{Department of Electrical and Computer Engineering, Rice University, Houston, Texas 77005, USA}
\affiliation{Department of Physics, National University of Singapore, Singapore 117551, Singapore}

\author{Oliver S.\ Dewey}
\affiliation{Carbon Hub, Rice University, Houston, Texas 77005, USA}
\affiliation{Department of Chemical and Biomolecular Engineering, Rice University, Houston, Texas 77005, USA}

\author{Lauren W.\ Taylor}
\affiliation{Carbon Hub, Rice University, Houston, Texas 77005, USA}
\affiliation{Department of Chemical and Biomolecular Engineering, Rice University, Houston, Texas 77005, USA}
\affiliation{Department of Chemical and Biomolecular Engineering, The Ohio State University, Columbus, OH 43210}

\author{Ali Mojibpour}
\affiliation{Department of Electrical and Computer Engineering, Rice University, Houston, Texas 77005, USA}

\author{Yingru Song}
\affiliation{Carbon Hub, Rice University, Houston, Texas 77005, USA}
\affiliation{Department of Mechanical Engineering, Rice University, Houston, Texas 77005, USA}

\author{Geoff Wehmeyer}
\affiliation{Carbon Hub, Rice University, Houston, Texas 77005, USA}
\affiliation{Department of Mechanical Engineering, Rice University, Houston, Texas 77005, USA}
\affiliation{Smalley--Curl Institute, Rice University, Houston, Texas 77005, USA}

\author{Matteo Pasquali}
\affiliation{Carbon Hub, Rice University, Houston, Texas 77005, USA}
\affiliation{Department of Chemical and Biomolecular Engineering, Rice University, Houston, Texas 77005, USA}
\affiliation{Smalley--Curl Institute, Rice University, Houston, Texas 77005, USA}
\affiliation{Department of Materials Science and NanoEngineering, Rice University, Houston, Texas 77005, USA}
\affiliation{Department of Chemistry, Rice University, Houston, Texas 77005, USA}

\author{Matthew S.\ Foster}
\affiliation{Carbon Hub, Rice University, Houston, Texas 77005, USA}
\affiliation{Department of Physics and Astronomy, Rice University, Houston, Texas 77005, USA}
\affiliation{Smalley--Curl Institute, Rice University, Houston, Texas 77005, USA}

\author{Douglas Natelson}
\affiliation{Department of Electrical and Computer Engineering, Rice University, Houston, Texas 77005, USA}
\affiliation{Carbon Hub, Rice University, Houston, Texas 77005, USA}
\affiliation{Department of Physics and Astronomy, Rice University, Houston, Texas 77005, USA}
\affiliation{Smalley--Curl Institute, Rice University, Houston, Texas 77005, USA}
\affiliation{Department of Materials Science and NanoEngineering, Rice University, Houston, Texas 77005, USA}

\author{Junichiro Kono}
\affiliation{Department of Electrical and Computer Engineering, Rice University, Houston, Texas 77005, USA}
\affiliation{Carbon Hub, Rice University, Houston, Texas 77005, USA}
\affiliation{Department of Physics and Astronomy, Rice University, Houston, Texas 77005, USA}
\affiliation{Smalley--Curl Institute, Rice University, Houston, Texas 77005, USA}
\affiliation{Department of Materials Science and NanoEngineering, Rice University, Houston, Texas 77005, USA}

             
\begin{abstract}
We investigate quantum transport in aligned carbon nanotube (CNT) fibers fabricated via solution spinning, focusing on the roles of structural dimensionality and quantum interference effects. The fibers exhibit metallic behavior at high temperatures, with conductivity increasing monotonically as the temperature decreases from room temperature to $\sim$36\,K. Below this temperature, the conductivity gradually decreases with further cooling, 
signaling the onset of quantum conductance corrections associated with localization effects.
Magnetoconductance measurements in both parallel and perpendicular magnetic fields exhibit pronounced positive corrections at low temperatures, consistent with weak localization (WL). To determine the effective dimensionality of electron transport, we analyzed the data using WL models in 1D, 2D, and 3D geometries. We found that while the 2D model can reproduce the field dependence, it lacks physical meaning in the context of our fiber architecture and requires an unphysical scaling factor to match the experimental magnitude. By contrast, we developed a hybrid 3D+1D WL framework that quantitatively captures both the field and temperature dependences using realistic coherence lengths and a temperature-dependent crossover parameter. Although this combined model also employs a scaling factor for magnitude correction, it yields a satisfactory fit, reflecting the hierarchical structure of CNT fibers in which transport occurs through quasi-1D bundles embedded in a 3D network. Our results establish a physically grounded model of phase-coherent transport in macroscopic CNT assemblies, providing insights into enhancing conductivity for flexible, lightweight power transmission applications.
\end{abstract}


\maketitle


\section{\label{sec:level1}Introduction}

Carbon nanotubes (CNTs) possess exceptional electrical conductivity, mechanical strength, and thermal properties at the single-tube level~\cite{iijima1996,yu2000strength,park2004,NanotetAl12AM}. Translating these intrinsic qualities into macroscopic assemblies has long been challenged by disorder and poor alignment, which typically lead to electronic transport dominated by localized states and thermally activated hopping processes~\cite{kaiser2001,skakalova2006}. In such systems, quantum effects are often masked, and conduction becomes sensitive to defects and intertube barriers. 

In recent years, significant progress has been made in the production of CNT fibers, leading to remarkable improvements in their alignment, packing density, and macroscopic conductivity. Advances in solution spinning and related processing techniques have enabled the fabrication of fibers with room-temperature conductivities exceeding 10~MS/m, values approaching or even surpassing those of some metals while maintaining mechanical robustness and low density~\cite{Behabtu2013,TAYLOR2021,dini2020}. These developments have been reviewed extensively, for example by Elliott and co-workers~\cite{elliott2021}, who summarized advances in processing strategies, doping methods, and the resulting structure--property relationships. Unlike randomly oriented CNT networks, which typically exhibit insulating-like temperature dependence with conductivity increasing upon heating~\cite{kaiser2001,skakalova2006}, highly aligned solution-spun CNT fibers often show metallic behavior over a wide temperature range~\cite{Behabtu2013,Piraux2015}. Despite these impressive achievements, most prior studies have emphasized enhancing bulk conductivity and mechanical performance, with comparatively little attention given to the fundamental transport mechanisms at the mesoscopic scale. In particular, the persistence of quantum interference effects and phase coherence in macroscopic CNT assemblies remains largely unexplored, motivating the present work.

Here, we studied CNT fibers produced by the solution spinning method~\cite{Behabtu2013}, which achieved a high degree of alignment and packing density. The specific fibers used in this work, commercially available and characterized in detail in Ref.\,\cite{TAYLOR2021}, exhibited among the highest reported room-temperature conductivities (10.9\,MS/m) for CNT macroscopic materials, while retaining excellent mechanical robustness. These fibers were among the first to reach a level of structural order sufficient for metallic conduction with quantum corrections. The suppression of disorder in such well-aligned structures allowed electron transport to transition from incoherent hopping to a regime where quantum interference becomes observable. Specifically, the temperature and magnetic field dependence of the conductance revealed clear signatures of weak localization (WL)—a hallmark of phase-coherent diffusive transport. At low temperatures, the magnetoconductance exhibited a pronounced positive correction that sharpened with decreasing temperature, consistent with time-reversal symmetry breaking and coherent backscattering in a metallic regime~\cite{bergmann1984,Altshuler1985,lin2002}.

However, a detailed analysis revealed that standard homogeneous 3D WL models could not capture the full behavior of the data. Instead, the experimental trends were more accurately described by treating the fiber as an ensemble of parallel quasi-one-dimensional CNT bundles, where the effective dimensionality of the system increased with decreasing temperature due to enhanced interbundle coupling. A numerical scaling factor was required to reconcile the model with the observed magnitude of the quantum correction, suggesting that anisotropic carrier diffusion and interbundle scattering played essential roles.

These results indicate that a full description of quantum transport in CNT fibers must incorporate both the internal coherence within individual bundles and the inelastic scattering processes governing charge transfer between bundles. Our findings provide a framework for understanding the interplay between dimensionality, disorder, and phase coherence in complex nanocarbon systems and point the way toward optimizing macroscopic CNT conductors for electronic applications.

\vspace{-12pt}
\section{\label{sec:level1}Sample and Experimental Methods}
\subsection{Sample Preparation}

CNT fibers were fabricated using the solution spinning method, detailed in references~\cite{Behabtu2013,TAYLOR2021}, with raw CNTs, primarily double-wall CNTs (DWCNTs), sourced from Meijo Nano Carbon Co. High-resolution transmission electron microscopy (HR-TEM) measurements yielded an average outer-wall diameter of $1.5$~nm and an average wall count of approximately 1.5. The viscosity-averaged aspect ratio was determined to be $4000$ via a capillary thinning extensional rheometer~\cite{Tsentalovich2016}, reflecting a high length-to-diameter ratio. The spinning solution, prepared with $0.5$~wt\% chlorosulfonic acid (CSA) (Sigma-Aldrich, $99$\%), was coagulated into acetone, forming fibers which were then collected on a rotating drum.

To produce further doped fibers for electrical transport measurements, we subjected the as-spun fibers to vapor-phase doping with iodine monochloride (ICl). The fibers were sealed in a nitrogen-purged round-bottom flask along with solid ICl and heated at $160^\circ$C in a box oven to facilitate dopant evaporation and diffusion. After doping, the flask was sequentially cooled in a water bath and then in an ice bath to promote dopant condensation and crystallization on the fiber surface.

\begin{figure*}[!ht]
\centering
\includegraphics[width=\textwidth]{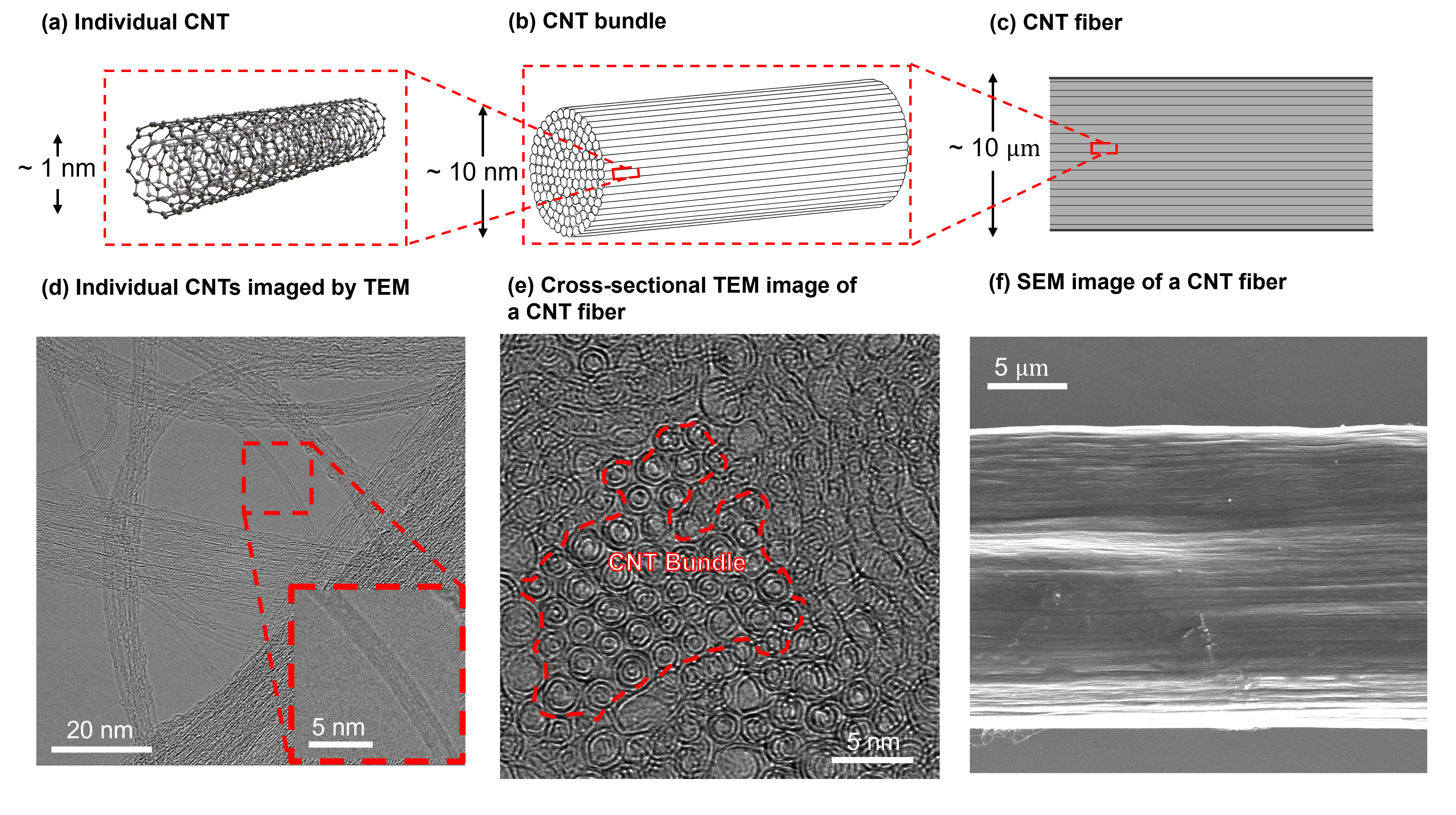}
\caption{
Hierarchical structure of CNT fibers. 
(a–c)~Schematic illustration of the multiscale organization formed during spinning: (a)~individual CNTs, (b)~aggregation into bundles  (each slender cylinder represents one CNT), and (c)~alignment of bundles into a macroscopic fiber (each elongated element denotes one bundle). 
(d–f)~Corresponding microscopy images: (d)~TEM image of raw CNTs before fiber spinning (the inset shows a zoomed-in region illustrating the double-walled tubular structure), (e)~cross-sectional TEM image showing densely packed bundles, and (f)~SEM image of the fiber surface exhibiting high alignment. 
Together, these images demonstrate how van der Waals interactions and flow-induced alignment give rise to long-range structural order across scales.
}
\label{fig:tube_bundle_fiber}
\end{figure*}

Figure~\ref{fig:tube_bundle_fiber}(a--c) schematically illustrates the internal multiscale structure formed during the spinning process. Individual CNTs, shown in Fig.~\ref{fig:tube_bundle_fiber}(a), spontaneously aggregate into bundles [Fig.~\ref{fig:tube_bundle_fiber}(b)], where each slender cylinder represents a single nanotube. These bundles, typically with diameters ranging from 10~nm to 100~nm, are densely packed and aligned within the fiber [Fig.~\ref{fig:tube_bundle_fiber}(c)], with each elongated element corresponding to one CNT bundle. This morphology results from collective interactions and flow-induced alignment during solution spinning, leading to an anisotropic structure with long-range order.

This schematic representation is supported by microscopy images in Fig.~\ref{fig:tube_bundle_fiber}(d--f). The transmission electron microscopy (TEM) image in Fig.~\ref{fig:tube_bundle_fiber}(d) shows individual CNTs prior to fiber spinning, exhibiting relatively straight segments and clean surfaces, indicating their suitability for tight bundling. Figure~\ref{fig:tube_bundle_fiber}(e) displays a cross-sectional TEM image of a CNT fiber, revealing distinct bundle domains and efficient packing across the fiber cross section. The scanning electron microscopy (SEM) image in Fig.~\ref{fig:tube_bundle_fiber}(f) shows the fiber surface, confirming a well-aligned morphology with an average diameter of 20.5\,$\upmu$m. Minor variations in fiber diameter observed along the axial direction are most likely the result of post-drawing relaxation or natural fluctuations formed during spinning. Similar diameter fluctuations have been noted in prior studies of highly aligned CNT fibers~\cite{Behabtu2013}, and are not necessarily indicative of structural inhomogeneity. The overall level of order and alignment observed here provides a reliable structural basis for the quantum transport phenomena discussed in the following sections.

\vspace{-22pt}

\subsection{Optical Measurements}

To elucidate the Fermi energy position ($E_\textrm{F}$) in ICl-doped CNT fibers, we conducted optical spectroscopy analysis on aligned CNT thin films fabricated by the blade-coating technique~\cite{HeadricketAl18AM}. CNTs were dissolved in CSA at a $0.5$~wt\% concentration, identical to that used in the fiber spinning process, to ensure consistency between the electronic states of the films and fibers. Direct optical spectroscopy on individual CNT fibers was not feasible due to their small diameter (10–20~$\upmu$m), which is significantly smaller than the typical Fourier transform infrared spectroscopy (FTIR) beam size ($\sim$1\,mm), resulting in insufficient signal-to-noise ratio. The solution was spread between two glass microscope slides mounted in a custom polytetrafluoroethylene (PTFE) holder. A PTFE stick was used to rapidly shear the slides apart at a rate of $\sim10^4$~s$^{-1}$, aligning the CNTs uniaxially. The sheared films were coagulated in acetone to remove CSA and retain the aligned structure. These as-formed films were then subjected to vapor-phase ICl doping, using the same process as that applied to the fibers.

Optical absorption spectra of CNT films were acquired using a custom optical measurement system, composed of a tungsten-halogen source (SLS201L, Thorlabs), a Glan-Thompson polarizer, and dual spectrometers for the spectral ranges of 550-1080\,nm and 1080-1425\,nm, respectively. For the spectral domain of 1425-3300\,nm, a Fourier transform infrared spectrometer (Nicolet™ iS50 FTIR Spectrometer, Thermo Fisher Scientific) was employed. The beam diameter was maintained at approximately 1~mm for measurements with both the custom setup and the Fourier transform infrared spectrometer.

\vspace{-16pt}
\subsection{Electronic Transport Measurements}
Electronic transport measurements were conducted using a magneto-optical cryostat (Oxford Spectromag-10T system) equipped with a variable temperature insert, utilizing the four-probe technique with gold wires and indium paste on a sapphire substrate for electrical contacts; the separation between inner probes was approximately 5 mm. Electrical resistance was measured across temperatures from 1.4 to 250\,K using dual lock-in amplifiers, maintaining the source current below 500\,$\upmu$A to mitigate sample heating, with Joule heating noted to be resistance-dependent. Magnetic fields up to 8~T, both perpendicular and parallel to the CNT fiber axis, were applied during transport measurements at selected temperatures (1.4, 10, 25, 50\,K), with electrical conductivity calculated from resistance, probe distance, and fiber diameter.

\vspace{-12pt}

\section{Experimental Results}
\subsection{\label{sec:level1}Temperature Dependence}

\begin{figure}[htb]
    \centering
    \includegraphics[width=0.48\textwidth]{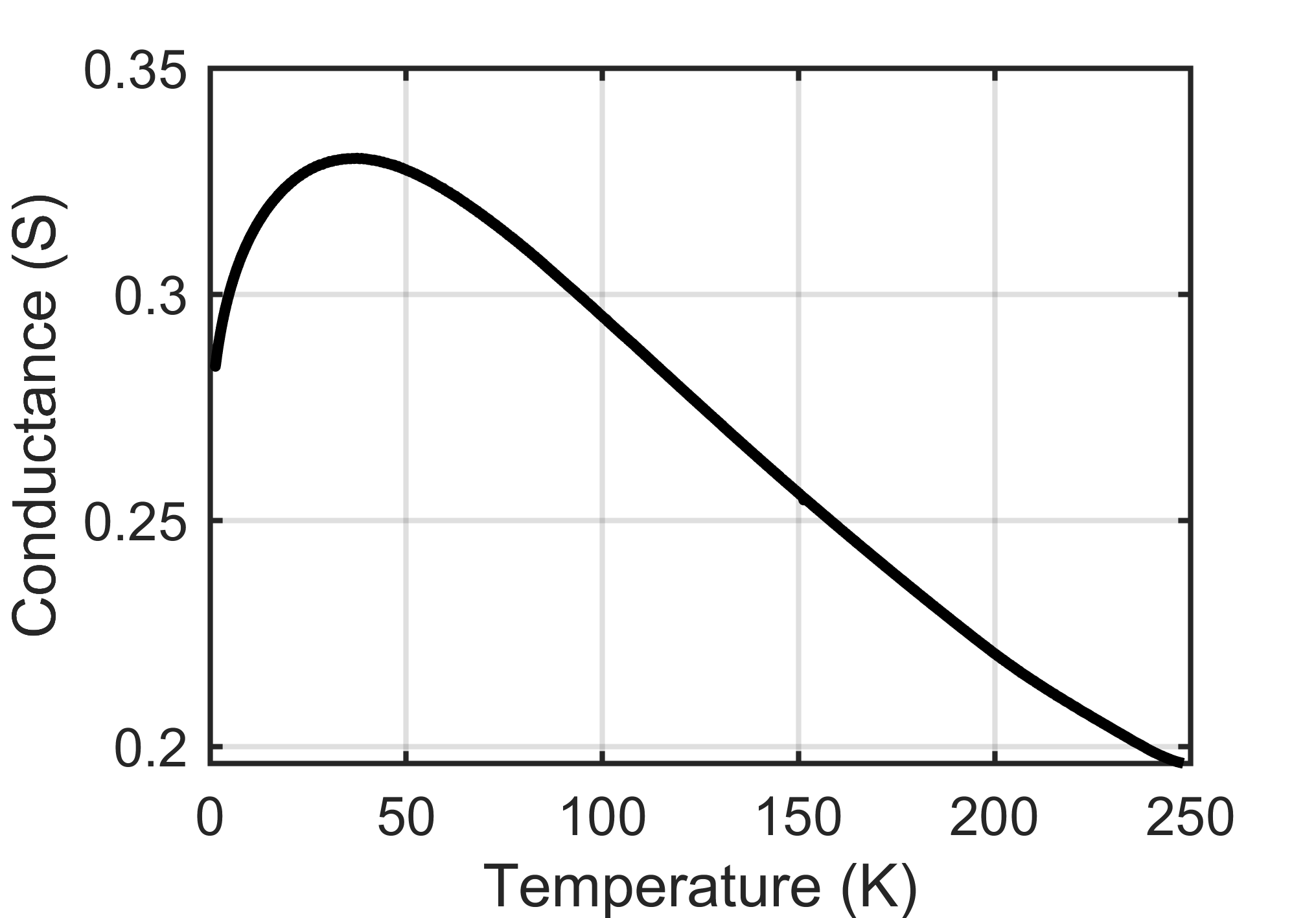}
    \caption{Temperature dependence of the electrical conductance for the ICl-doped fiber sample. The conductance increases upon cooling, reaches a maximum near 36\,K, and decreases at lower temperatures. This nonmonotonic behavior is consistent with previous reports on doped carbon nanotube materials~\cite{Behabtu2013}. The curve shows densely sampled experimental data points directly connected without smoothing.}
    \label{fig:Conductance_vs_Temperature}
\end{figure}

Figure~\ref{fig:Conductance_vs_Temperature} shows the temperature dependence of electrical conductance for the ICl-doped fiber sample. The conductance exhibits a nonmonotonic behavior: it increases with decreasing temperature, reaches a maximum at approximately 36~K, and then decreases at lower temperatures. This behavior is consistent with previous studies on high quality, high conductivity doped carbon nanotube fibers~\cite{Behabtu2013}. 

\subsection{\label{sec:level1}Magnetoconductance and Weak Localization Modeling}

\begin{figure}
    \centering
    \begin{minipage}[t]{\linewidth}
        \centering
        \begin{overpic}[width=\linewidth]{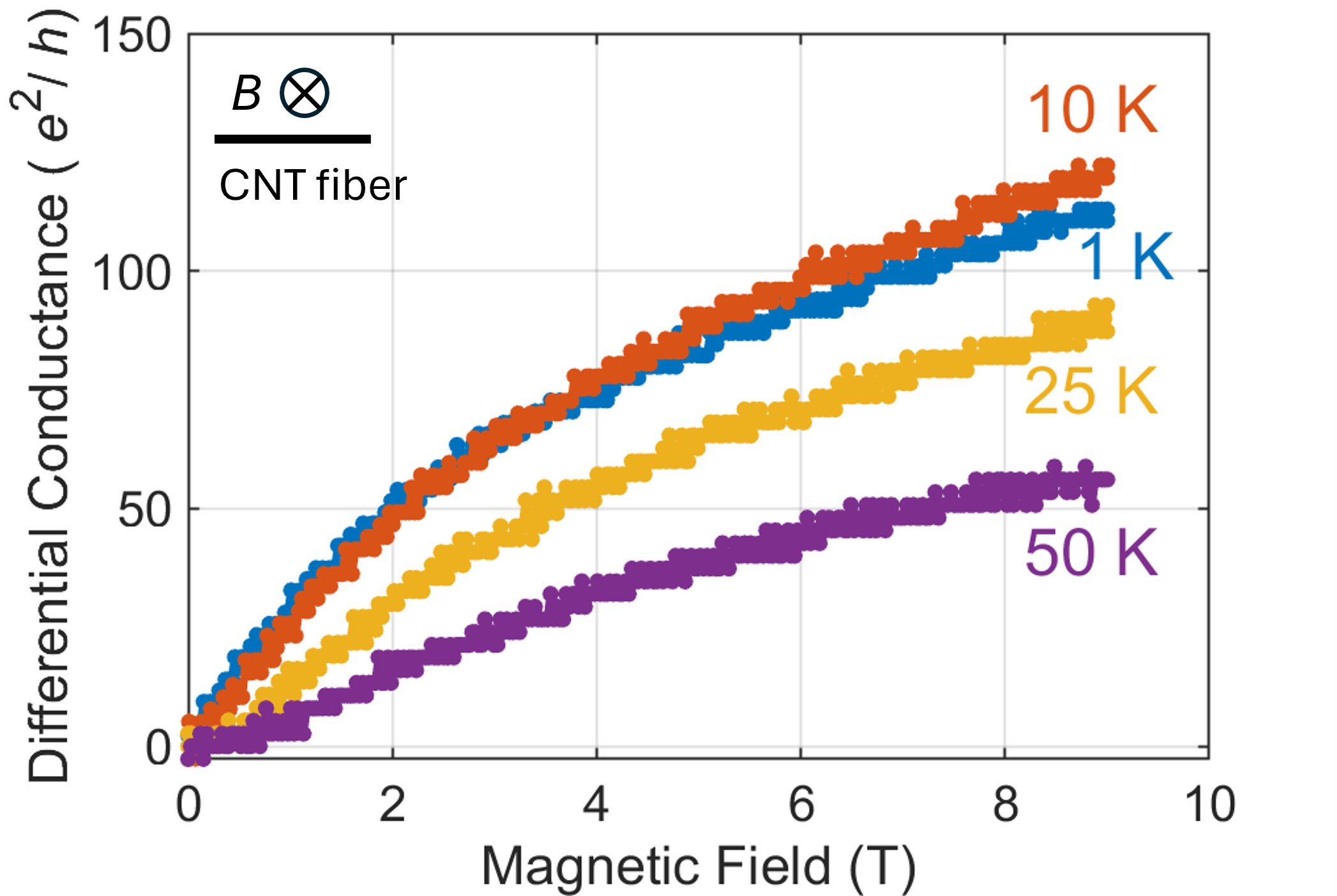}
            \put(5,70){\textbf{\textsf{(a) Perpendicular}}} 
        \end{overpic}
    \end{minipage}
    \vspace*{2mm}
    
    \begin{minipage}[t]{\linewidth}
        \centering
        \begin{overpic}[width=\linewidth]{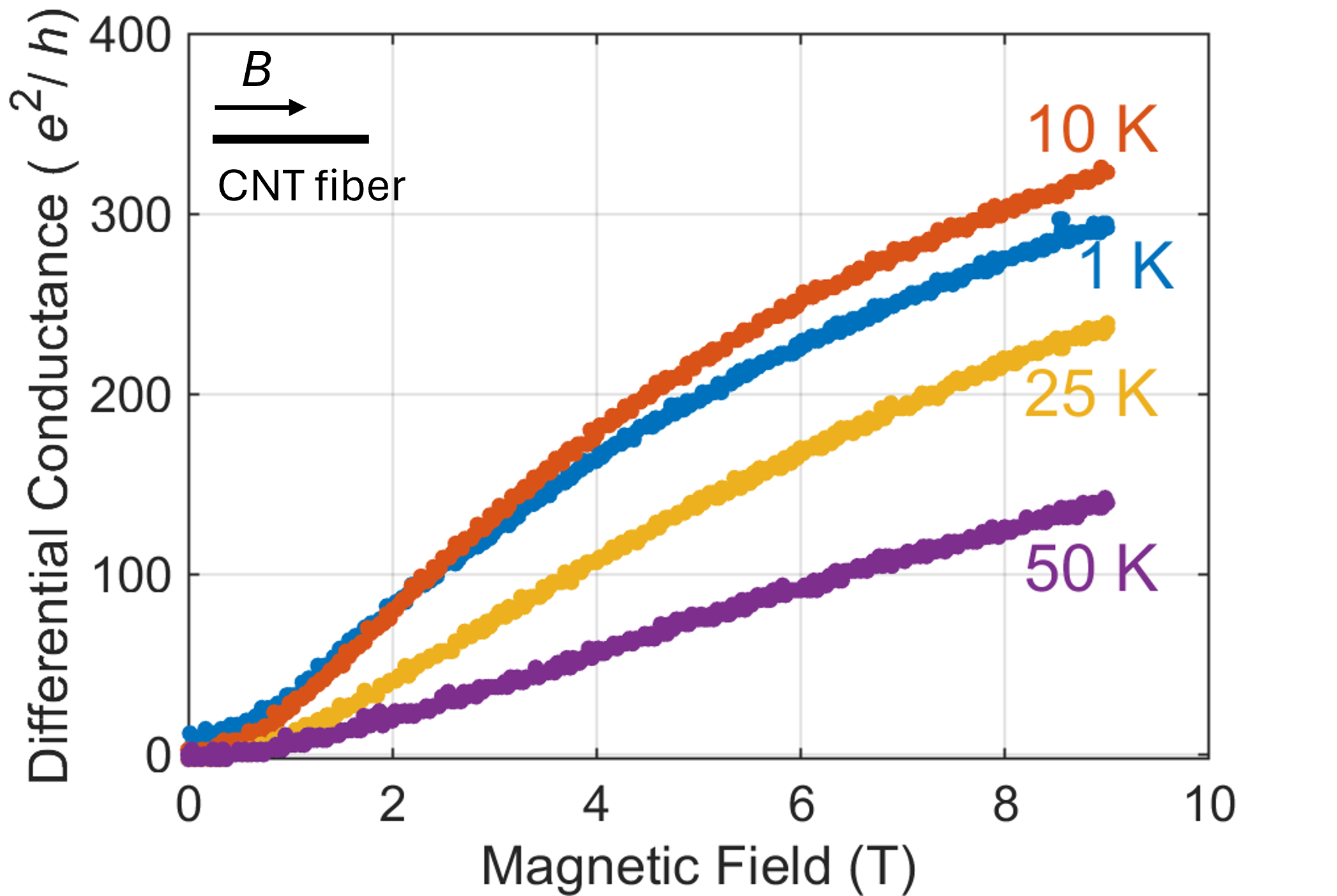}
            \put(5,70){\textbf{\textsf{(b) Parallel}}} 
        \end{overpic}
    \end{minipage}

    \caption{
        (Color online) Differential conductance as a function of magnetic field for (a) perpendicular and (b) parallel orientations, measured at various temperatures. The conductance is expressed in units of \( e^2/h \). Data were taken at \( T = 1~\mathrm{K} \) (blue), \( 10~\mathrm{K} \) (red), \( 25~\mathrm{K} \) (yellow), and \( 50~\mathrm{K} \) (purple).
    }
    \label{fig:DiffConductance}
\end{figure}

In a disordered, diffusive conductor, WL originates from constructive interference of time-reversed electron trajectories on self-intersecting loops, which enhances backscattering and reduces the zero-field conductance. An applied magnetic field breaks the time-reversal symmetry of these loops, suppressing the WL correction and thereby yielding a positive magnetoconductance (MC). At lower temperatures, reduced inelastic scattering increases the phase-coherence length $L_\varphi$, strengthening the WL correction and producing a conductance downturn. 

Two primary mechanisms can produce positive MC: WL and forward interference. However, forward interference is typically relevant in the insulating regime governed by variable-range hopping (VRH), where it manifests as orbital MC; a classic example is the model combining forward interference and wave-function shrinkage in Mott VRH systems~\cite{Rosenbaum2001}. As our system is metallic and not in the VRH regime, WL---not forward interference---is the appropriate framework for modeling our data.

Figure~\ref{fig:DiffConductance} shows the differential conductance change, defined as \(\Delta G(B) = G(B) - G(0)\), as a function of magnetic field for the ICl-doped fiber sample at different temperatures, under (a) perpendicular and (b) parallel magnetic field orientations. 
In both configurations, the conductance increases with increasing magnetic field, indicative of the suppression of quantum interference effects such as weak localization. The parallel field configuration exhibits a significantly stronger field dependence, suggesting enhanced phase coherence or reduced dimensionality along the nanotube alignment direction. The amplitude of \(\Delta G(B)\) also grows at lower temperatures, which is consistent with the increasing coherence length. These trends point to the presence of phase-coherent transport and magnetic-field-induced delocalization in the system.

To quantitatively analyze the magnetoconductance, we employed WL theory in diffusive systems. The conductance correction due to WL arises from quantum interference between time-reversed electron trajectories and is suppressed by an external magnetic field. The effective dimensionality of the system determines the functional form of the field dependence.  In this way, such magnetotransport measurements are sensitive to whether carriers go from tube to tube and bundle to bundle via elastic or inelastic processes. Since WL theories are valid in the low-field regime where quantum interference dominates, our fitting analysis is therefore restricted to fields below approximately 1\,T.

The effective dimensionality of WL is governed by the ratio between the electronic coherence length, $L_\varphi$, and the sample dimensions. When $L_\varphi$ is smaller than the sample length $L$, width $w$, and thickness $t$, the system falls in the quasi-3D regime. If the width and thickness are both smaller than $L_\varphi$, but $L_\varphi$ remains shorter than $L$, the behavior is quasi-2D. Conversely, when the thickness is smaller than $L_\varphi$, which in turn is shorter than both $L$ and $w$, the system lies in the quasi-1D regime. Importantly, the WL correction to the conductance takes on different functional forms in each case, and crossovers between these regimes occur gradually, rather than abruptly, as $L_\varphi$ evolves with temperature.

\subsubsection{1D Model Fit}
\label{sec:1D_fit}

To facilitate direct comparison between quasi-one-dimensional (1D) WL theory and experimental measurements on macroscopic CNT fibers, we construct an effective conductivity model by rescaling the standard 1D WL expression.

We begin with the quasi-1D WL conductivity correction for a single nanotube bundle in a perpendicular magnetic field~\cite{Beenakker1991, Altshuler1985}:
\begin{equation}
\delta \sigma_{\mathrm{1D},\perp}(B) = -\frac{2e^2}{h} 
 \frac{1}{\sqrt{L_{\varphi, \mathrm{1D}}^{-2} + \dfrac{A_\text{b}}{3\ell_B^4}}},
\label{eq:quasi1D}
\end{equation}
where \( A_\text{b} \) is the effective cross-sectional area of a bundle, \( \ell_B = \sqrt{\hbar/(eB)} \) is the magnetic length, and \( L_\varphi \) is the phase coherence length.

To extend this result to the macroscopic fiber, we model the system as comprising \( N \) individual parallel bundles, each contributing a conductance \( G_\text{b} \), such that the total conductance is \( G_N = N G_\text{b} \). The corresponding three-dimensional (3D) conductivity is given by
\begin{equation}
\sigma_{\mathrm{3D}} = \left( \frac{N}{A_\text{F}} \right) \sigma_{\mathrm{1D}},
\label{eq:sigma3D_N}
\end{equation}
where \( A_\text{F} \) is the cross-sectional area of the fiber and \( \sigma_{\mathrm{1D}} \) is the conductivity of a single bundle. Assuming the bundles are closely packed and have uniform cross-sectional area \( A_\text{b} \), we approximate the bundle density as \( N / A_\text{F} \approx 1 / A_\text{b} \), yielding
\begin{equation}
\sigma_{\mathrm{3D}} = \frac{1}{A_\text{b}} \sigma_{\mathrm{1D}}.
\label{eq:sigma3D}
\end{equation}

Substituting Eq.\,\eqref{eq:quasi1D} into Eq.\,\eqref{eq:sigma3D}, we obtain the magnetic-field-dependent conductivity correction:
\begin{equation}
\delta \sigma_{\mathrm{3D}}(B) = \frac{1}{A_\text{b}} \delta \sigma_{\mathrm{1D}}(B) 
= -\frac{2e^2}{h A_\text{b}} \frac{1}{\sqrt{L_{\varphi,\mathrm{1D}}^{-2} + \dfrac{A_\text{b}}{3\ell_B^4}}}.
\label{eq:sigma3D_final}
\end{equation}
This expression shows how quasi-1D WL in individual bundles contributes to the effective 3D conductivity behavior of the macroscopic fiber.

To relate the conductivity change to the experimentally measured conductance difference, we apply a geometric prefactor:
\begin{equation}
\Delta G(B) = \frac{A_\text{F}}{L} \left[ \sigma(B) - \sigma(0) \right] = \frac{A_\text{F}}{L} \left[ \delta\sigma(B) - \delta\sigma(0) \right],
\label{eq:conductance_conversion}
\end{equation}
where \( L \) is the channel length. Combining Eqs.\,\eqref{eq:sigma3D_final} and \eqref{eq:conductance_conversion}, we obtain the final expression for the conductance change under magnetic field:
\begin{equation}
\Delta G(B) = 
-\frac{2 A_\text{F} e^2}{A_\text{b} L h} 
\left[ 
\frac{1}{\sqrt{L_{\varphi,\mathrm{1D}}^{-2} + \dfrac{A_\text{b}}{3 \ell_B^4}}}
- L_{\varphi,\mathrm{1D}} 
\right],
\label{eq:1D_final}
\end{equation}
where \(\Delta G(B)\) is the magnetoconductance. All other symbols are as previously defined.

\begin{figure}[htb]
    \centering
    \includegraphics[width=0.45\textwidth]{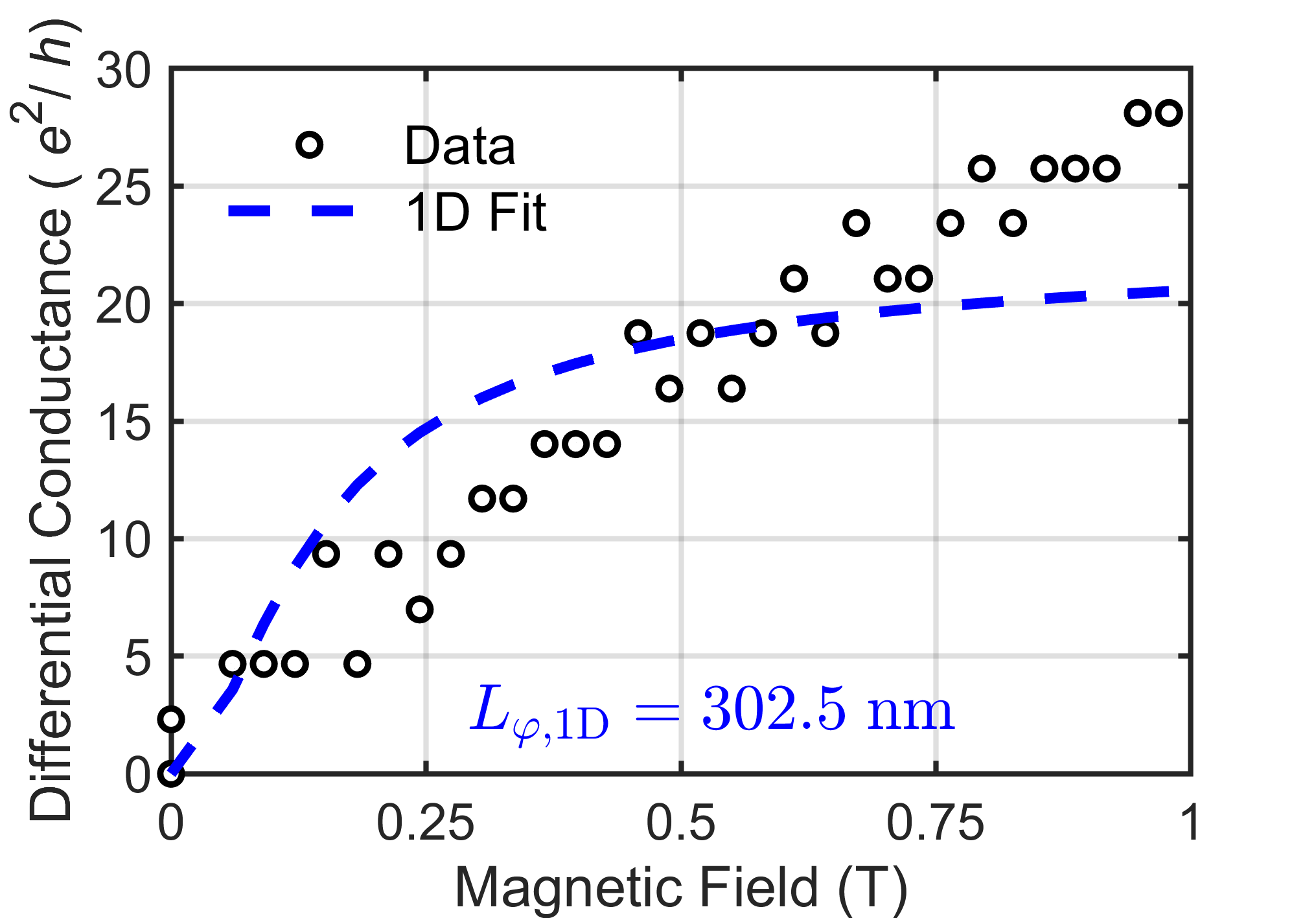}
    \caption{
        Fit of the 1D WL model to the measured \(\Delta G(B)\) at 1\,K. The model captures the overall scale of the conductance correction but fails to reproduce the detailed field dependence, overestimating the curvature at low fields. The bundle cross-sectional area used in the fit is \(A_\text{b} = (40~\mathrm{nm})^2\).
        }
    \label{fig:1D_fit}
\end{figure}

This expression was used to fit the experimental \(\Delta G(B)\) data at 1\,K for the perpendicular field configuration, treating the 1D coherence length \( L_{\varphi,\mathrm{1D}} \) as a free parameter. Figure~\ref{fig:1D_fit} shows the resulting fit. While the overall magnitude of the conductance correction is reasonably captured, the theoretical curve fails to reproduce the detailed magnetic field dependence at low \( B \). In particular, it overestimates the curvature in the low-field regime, rising more sharply than the experimental data. This discrepancy suggests that the effective dimensionality of the system may deviate from the strictly 1D limit.

\vspace{16pt}
\subsubsection{3D Model Fit}
\label{sec:3D_fit}

To provide a baseline for comparison, we also consider the isotropic 3D WL model~\cite{Altshuler1985}.  This is equivalent to assuming that the fiber is a homogeneous conductor and that the coherence length is smaller than both the wire diameter and its length. The 3D conductivity correction in this case can be written as:
\begin{equation}
\delta \sigma_\text{3D}(B) = -\frac{e^2}{\pi h} \left( \frac{1}{\ell_\text{el}} - \frac{1}{L_{\varphi, \text{3D}}} \right) 
+ \frac{e^2}{\pi h \ell_B} f_3\left( \frac{4L_{\varphi, \text{3D}}^2}{\ell_B^2} \right),
\label{eq:delta_sigma_3D}
\end{equation}
where \( \ell_\text{el} \) is the elastic mean free path, and \( L_{\varphi, 3D} \) is the phase coherence length. We employ the full numerical form of the function \( f_3(x) \), evaluated as
\begin{multline}
f_3(x) = \sum_{n=0}^{\infty} \Bigg[
  2\sqrt{n + \frac{1}{x} + 1}
  - 2\sqrt{n + \frac{1}{x}} \\
  - \left(n + \frac{1}{x} + \frac{1}{2} \right)^{-1/2}
\Bigg],
\label{eq:f3_sum}
\end{multline}
with the sum truncated numerically at a large enough \( n \) to ensure convergence~\cite{Kawabata1980}.
Its asymptotic behavior is
\begin{equation}
f_3(x) \approx 
\begin{cases}
\displaystyle \frac{x^{3/2}}{48}, & x \ll 1 \quad \text{(weak-field limit)}, \\
0.605, & x \gg 1 \quad \text{(strong-field limit)}.
\end{cases}
\label{eq:f3_asymptotics}
\end{equation}
While this model assumes isotropic 3D transport, it provides a useful point of comparison for identifying dimensional crossover behavior in anisotropic or quasi-1D systems such as aligned CNT fibers.

To perform the 3D fit, we used the following expression:
\begin{align}
    \Delta G(B) 
    &= \frac{A_\text{F}}{L} \left[ \delta\sigma(B) - \delta\sigma(0) \right] 
\nonumber\\        
    &= \frac{e^2}{h} \frac{A_\text{F}}{L} \frac{1}{\pi \ell_B} 
       f_3\left( \frac{4 L_{\varphi, 3\text{D}}^2}{\ell_B^2} \right).
\label{eq:3D_WL}
\end{align}
Although Eq.\,(\ref{eq:3D_WL}) captures the characteristic field dependence of 3D WL, it yields only a limited conductance correction. For representative parameters (fiber width 20\,\(\upmu\)m, length 5\,mm), the geometric prefactor \( A_\text{F}/(\pi L \ell_B) \) evaluates to approximately 0.77 at \( B = 1~\mathrm{T} \). Moreover, the function \( f_3(x) \) saturates at 0.6 for \( x \gg 1 \), so even in the most favorable regime allowed by the model, the predicted \(\Delta G(B)\) remains well below the measured values.

This analysis demonstrates that the 3D model significantly underestimates the quantum correction and is therefore inadequate to describe the transport behavior in our system. These results support a lower-dimensional interpretation, consistent with the anisotropic or quasi-1D structure of aligned CNT fibers.

\subsubsection{2D Model Fit}
To explore whether a two-dimensional (2D) WL framework could describe the observed magnetoconductance behavior, we applied the model used in Ref.\,\cite{Piraux2015,sciRep2017}, expressed as
\begin{widetext}
\begin{equation}
\Delta G(B) = -\frac{e^2}{\pi h} N_{\text{LAYER}} 
\left\{
\ln\left[\left(\frac{1}{L_{\varphi, \text{2D}}^{2}}\right) \frac{\hbar}{4eB} \right]
- \psi\left(\frac{1}{2} + \left(\frac{1}{L_{\varphi, \text{2D}}^{2}}\right) \frac{\hbar}{4eB} \right)
\right\}
\label{eq:2D_WL}
\end{equation}
\end{widetext}
where \( \psi \) is the digamma function, and \( N_{\text{LAYER}} \) is an empirical scale factor introduced to match the magnitude of the experimental data.

\begin{figure}[htb]
    \centering
    \includegraphics[width=0.5\textwidth]{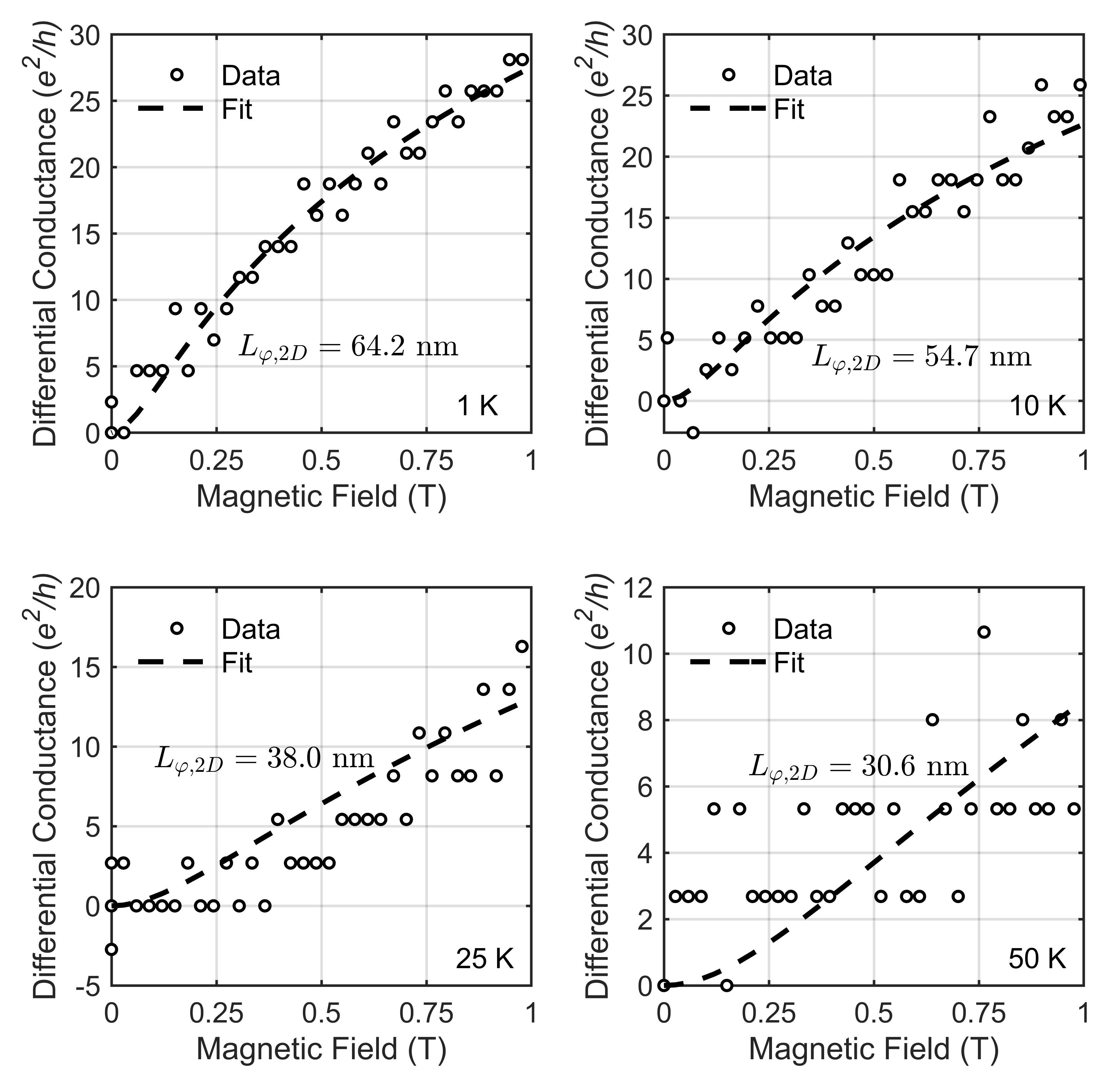}
    \caption{
    Two-dimensional weak localization fits to the differential conductance data under perpendicular magnetic fields at four representative temperatures: 1\,K, 10\,K, 25\,K, and 50\,K. 
    The experimental data (circles) are fitted using the model from Ref.\,\onlinecite{Piraux2015}, as given in Eq.\,\eqref{eq:2D_WL}, with \( L_{\varphi, 2\text{D}} \) as the sole free parameter.
    While the fits qualitatively reproduce the field dependence, they require a prefactor \( N_{\text{LAYER}} \) to match the magnitude of the measured \(\Delta G\), and the extracted coherence lengths are shown in each panel.
    }
    \label{fig:2D_fit}
\end{figure}

While this model reproduces the overall curvature of the \( \Delta G(B) \) traces, as shown in Fig.~\ref{fig:2D_fit}, the fit requires a prefactor \( N_{\text{LAYER}} \approx 60 \), serving as a scale factor. The need for such an empirical scaling term suggests that the standard 2D expression does not adequately capture the magnitude of the conductance correction in our system, likely due to a mismatch between the model dimensionality and the actual transport geometry. Namely, this adjustment lacks a clear microscopic basis, and \( N_{\text{LAYER}} \) does not correspond to any well-defined physical layer count in our sample.

Moreover, the use of a 2D WL model is not physically justified for our sample geometry. In localization theory, the effective dimensionality is governed by the relation between the phase coherence length \( L_\varphi \) and the characteristic sample dimensions. As illustrated in Fig.\,\ref{fig:tube_bundle_fiber}, 
our system exhibits three relevant structural length scales: the fiber diameter (\( 20.5\,\upmu\text{m} \)), the bundle diameters  (\( \sim\!10\text{\,--\,}100\,\text{nm} \)), and the diameters of individual nanotubes (\( \sim 1.5\,\text{nm} \)). A 2D regime typically assumes that \( L_\varphi \) exceeds the sample thickness but remains smaller than the lateral width, conditions commonly satisfied in thin-film systems. However, the constituent elements of our samples—cylindrical tubes and bundles arranged in quasi-1D configurations — do not form an extended planar geometry. 

Therefore, despite the  agreement with the functional form of the experimental data, the 2D WL model cannot be considered a physically meaningful description of our system. The fitting parameter \( N_{\text{LAYER}} \) compensates for a mismatch in model dimensionality, and the extracted coherence lengths \( L_{\varphi, 2\text{D}} \) do not correspond to a physically interpretable spatial scale. We conclude that a dimensional crossover between 1D and 3D WL is physically more appropriate to describe transport in this system, consistent with the hierarchical structure of aligned carbon nanotube networks.

\vspace{16pt}
\subsubsection{Hybrid 1D+3D Model and Dimensional Crossover}

The 1D and 3D WL models, presented in Sections~\ref{sec:1D_fit} and \ref{sec:3D_fit}, describe phase-coherent transport along individual bundles and diffusive averaging across the entire fiber, respectively. However, as discussed therein, neither model alone adequately reproduces the experimental magnetoconductance. This motivates a hybrid approach in which the 1D WL correction is applied at the bundle scale, combined with a geometric rescaling appropriate for the macroscopic 3D structure.

This interpretation is supported by the hierarchical morphology of the CNT fibers (Fig.~\ref{fig:tube_bundle_fiber}) and by prior studies identifying bundles—not individual tubes—as the primary transport units~\cite{Piraux2015,Wang2018}. The SEM image in Fig.\,\ref{fig:tube_bundle_fiber}(f) highlights the prevalence of bundles as dominant structural elements, and the bundle size distribution shown in Fig.\,S2 further supports the applicability of a mixed model when the coherence length \( L_\varphi \sim 10~\text{nm} \) exceeds some bundles but remains smaller than others.

To capture the intermediate-dimensional nature of transport in aligned CNT fibers, we adopt a composite model that incorporates both quasi-1D and 3D WL contributions. The total quantum correction to the conductivity is modeled as a weighted sum of the 3D and 1D components:
\begin{equation}
    \delta \sigma(B) 
    = 
    M(T) 
    \cdot 
    \delta \sigma_\text{3D}(B) 
    + 
    [1 - M(T)] 
    \cdot 
    \frac{1}{A_\text{b}}
    \delta \sigma_\text{1D}(B),
    \label{eq:combined_sigma_pre}
\end{equation}
where \( M(T) \) is a phenomenological weighting factor that captures the temperature-dependent dimensional crossover, and each component is defined by Eqs.\,\eqref{eq:delta_sigma_3D} and \eqref{eq:quasi1D}.
Here $A_\text{b}$ is the typical bundle cross sectional area; see Eq.\,(\ref{eq:sigma3D_final}). The conductivity change induced by the magnetic field, $B$, is then given by
\begin{widetext}
\begin{equation}
\sigma(B) - \sigma(0) = \delta \sigma(B) - \delta \sigma(0)
= \frac{e^2}{h} \left\{
M(T) \cdot \frac{1}{\pi \ell_B} f_3\left[ \frac{4 L_{\varphi,\text{3D}}^2}{\ell_B^2} \right] 
- \left[ 1 - M(T) \right] \cdot \frac{2}{A_\text{b}} 
\left[ 
\frac{1}{\sqrt{L_{\varphi,\text{1D}}^{-2} + \frac{A_\text{b}}{3} \left( \frac{1}{\ell_B^4} \right)}}
- L_{\varphi,\text{1D}}
\right]
\right\}.
\label{eq:combined_sigma}
\end{equation}
\end{widetext}

To convert the conductivity difference into the experimentally measurable conductance difference, we multiply it by the fiber’s cross-sectional area \( A_\text{F} \) and divide it by its length \( L \):
\begin{equation}
G(B) - G(0) = \frac{A_\text{F}}{L} \left[ \sigma(B) - \sigma(0) \right].
\label{eq:combined_conductance}
\end{equation}
Here, \( L = 0.55\,\text{cm} \) is the fiber channel length, and \( A_\text{F} = \pi \times (10.25 \times 10^{-4}~\text{cm})^2 \) is the cross-sectional area of the fiber. 
The final expressions, 
Eqs.\,\eqref{eq:combined_sigma} and \eqref{eq:combined_conductance}, 
were used to fit the field-dependent conductance data with \( L_{\varphi,\text{1D}}(T) \), \( L_{\varphi,\text{3D}}(T) \), and \( M(T) \) as the fitting parameters.

We applied Eq.\,\eqref{eq:combined_conductance} to fit the magnetic field dependence of the differential conductance at four representative temperatures: 1\,K, 10\,K, 25\,K, and 50\,K. For each temperature, we treated the coherence lengths \( L_{\varphi,\text{1D}}(T) \) and \( L_{\varphi,\text{3D}}(T) \) as independent fitting parameters, along with the crossover weight \( M(T) \), which is labeled in each panel of the fit results. The function \( f_3(x) \) in the 3D contribution was scaled by a phenomenological scale factor of 200, 
consistent with previous studies where empirical magnitude correction was necessary to reconcile the WL theory with conductance measurements on CNT systems~\cite{Piraux2015}. 

The values of \( L_{\varphi,\text{1D}} \) and \( L_{\varphi,\text{3D}} \) obtained through the fitting procedure are summarized in Table~\ref{tab:Lphi_fits}, providing insights into the temperature dependence of the quantum coherence lengths along both dimensional components. As expected, both coherence lengths decrease with increasing temperature, consistent with decoherence from thermal fluctuations.

\begin{table}[h]
\centering
\caption{Extracted phase coherence lengths from the 3D+1D WL fits. 
A single crossover weight \( M(T) \) is adopted for each temperature, 
representing the approximate average across field orientations. 
}
\label{tab:Lphi_fits}
\begin{tabular}{c|c|c|c|c}
\hline
Temperature & Orientation & \( L_{\varphi,\text{3D}} \) (nm) & \( L_{\varphi,\text{1D}} \) (nm) & \( M \) \\
\hline
\multirow{2}{*}{1 K}  & \( \perp \)    & 140 & 48 & \multirow{2}{*}{\( \approx 0.40 \)} \\
                      & \( \parallel \) & 180 & 33 & \\
\hline
\multirow{2}{*}{10 K} & \( \perp \)    & 59  & 23 & \multirow{2}{*}{\( \approx 0.60 \)} \\
                      & \( \parallel \) & 54  & 22 & \\
\hline
\multirow{2}{*}{25 K} & \( \perp \)    & 25  & N/A & \multirow{2}{*}{\( \approx 1.00 \)} \\
                      & \( \parallel \) & 25  & N/A & \\
\hline
\multirow{2}{*}{50 K} & \( \perp \)    & 16  & N/A & \multirow{2}{*}{\( \approx 1.00 \)} \\
                      & \( \parallel \) & 15  & N/A & \\
\hline
\end{tabular}
\end{table}

\begin{figure}
    \centering
    \begin{minipage}[t]{\linewidth}
        \centering
        \begin{overpic}[width=\linewidth]{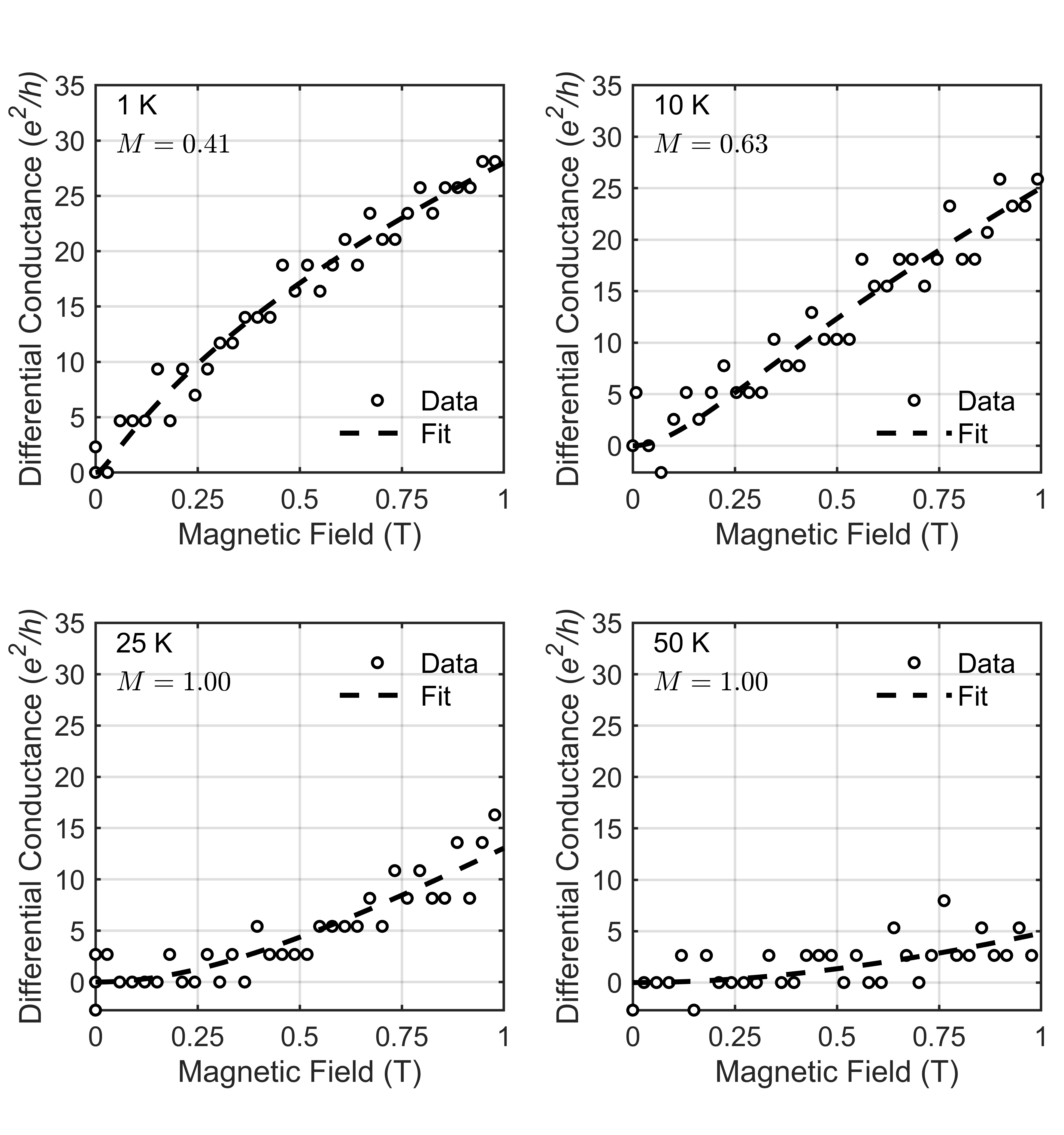}
            \put(5,96){\textbf{\textsf{(a) Perpendicular}}} 
        \end{overpic}
    \end{minipage}

    \begin{minipage}[t]{\linewidth}
        \centering
        \begin{overpic}[width=\linewidth]{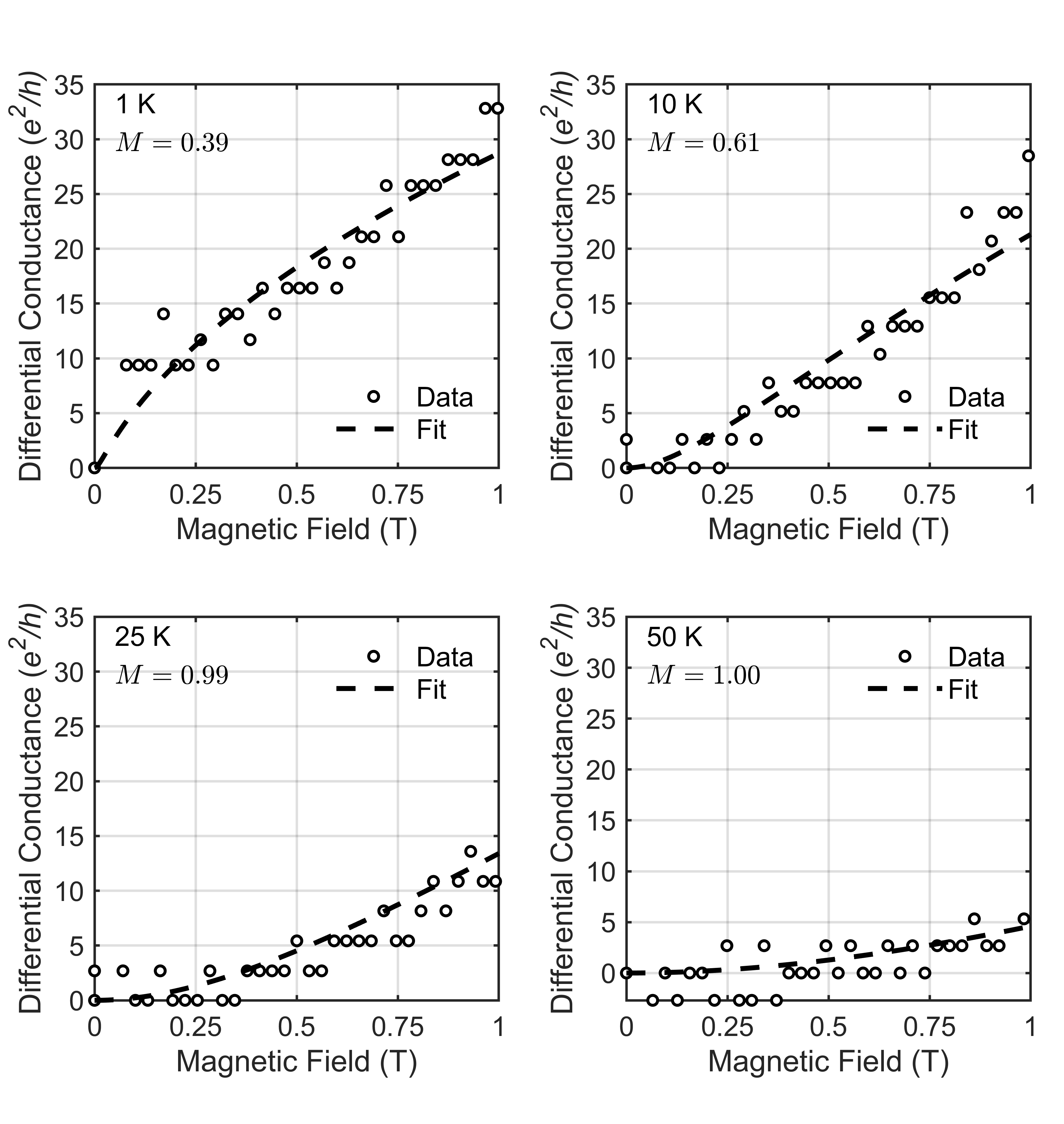}
            \put(5,97){\textbf{\textsf{(b) Parallel}}} 
        \end{overpic}
    \end{minipage}
    \caption{
Composite 3D+1D weak-localization fits to differential conductance under (a) perpendicular and (b) parallel magnetic fields at 1\,K, 10\,K, 25\,K, and 50\,K. Data (circles) and fits (dashed lines) are shown, with the crossover parameter \(M(T)\) labeled. A phenomenological scale factor of \(\sim 200\) was applied. All panels use a fixed y-axis limit of 35~\(e^{2}/h\); insets highlight the 25\,K and 50\,K data, where the magnetoconductance becomes weaker. Agreement across field orientations supports mixed-dimensional transport.
    }
    \label{fig:3D1D_fits_combined}
\end{figure}

Figure\,\ref{fig:3D1D_fits_combined}(a) shows the fitting results for the perpendicular field configuration. The model successfully captures the magnitude and curvature of the conductance across all temperatures. The crossover factor \( M \), labeled in each panel, increases with temperature, consistent with a gradual transition from 1D-dominated to 3D-dominated behavior. Similar trends are observed in the parallel field configuration [Fig.\,\ref{fig:3D1D_fits_combined}(b)], supporting the robustness of the mixed-dimensional WL interpretation.

\begin{figure}[htb]
    \centering
    \includegraphics[width=0.45\textwidth]{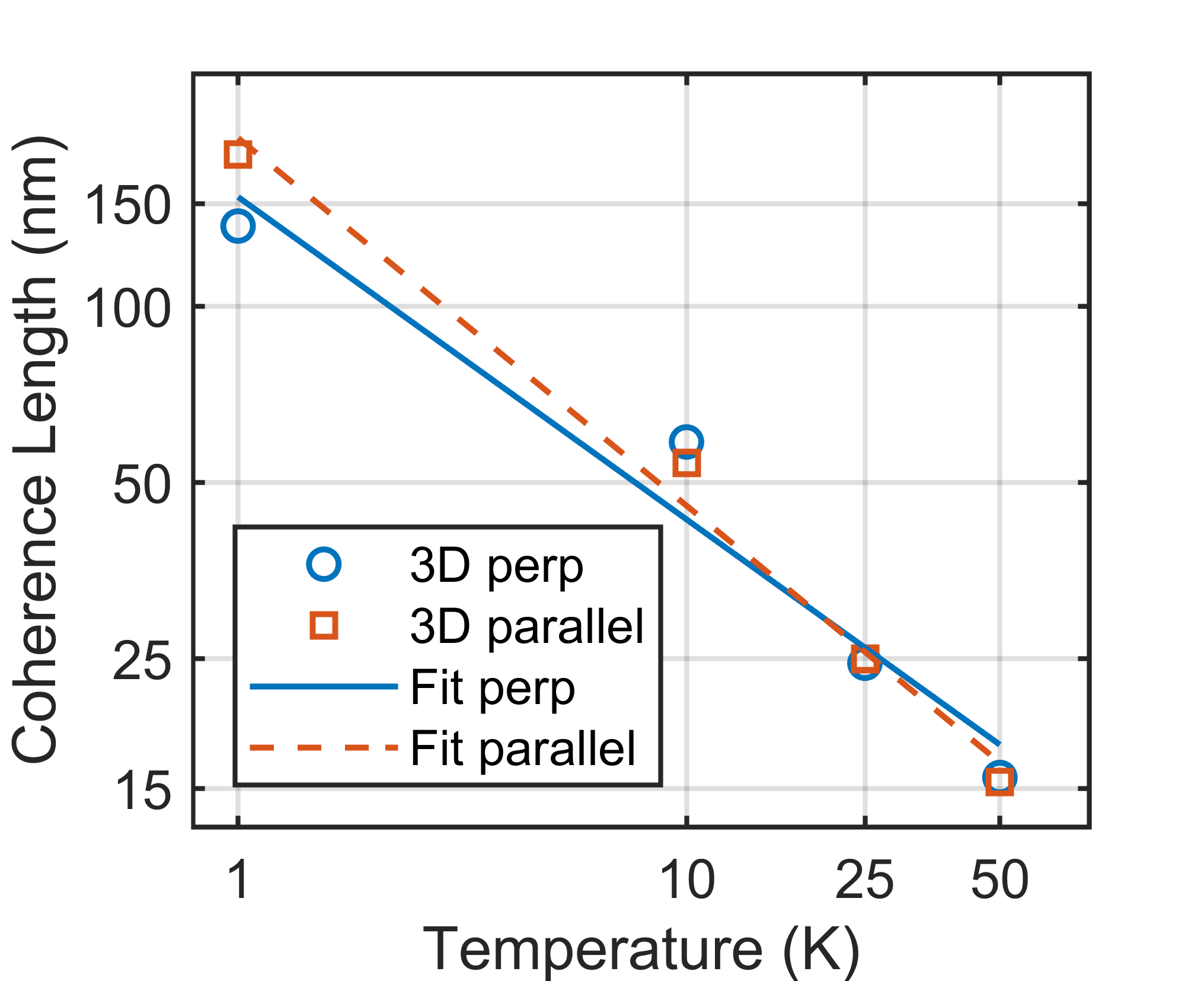}
    \caption{
    Temperature dependence of the 3D phase coherence length \( L_{\varphi, \text{3D}} \) extracted from weak localization fits for both perpendicular (blue circles) and parallel (red squares) magnetic fields. The solid and dashed lines represent power-law fits to the perpendicular and parallel data, respectively. A consistent decay of \( L_{\varphi, \text{3D}} \) with increasing temperature is observed, indicating enhanced decoherence at higher temperatures.
    }
    \label{fig:Lphi3D}
\end{figure}

The extracted 3D coherence lengths $L_{\varphi,3D}(T)$ decrease with temperature, and the log--log fits yield slopes of $-0.70$ (perpendicular field) and $-0.77$ (parallel field), both close to the theoretical WL exponent of $-0.75$ ($-3/2$) \cite{Altshuler1985,lin2002}, as summarized in Fig.~\ref{fig:Lphi3D}.

\vspace{16pt}
\subsection{High-Temperature Transport and Residual Scattering Mechanism}
\begin{figure}
\flushleft
\includegraphics[width=0.46\textwidth]{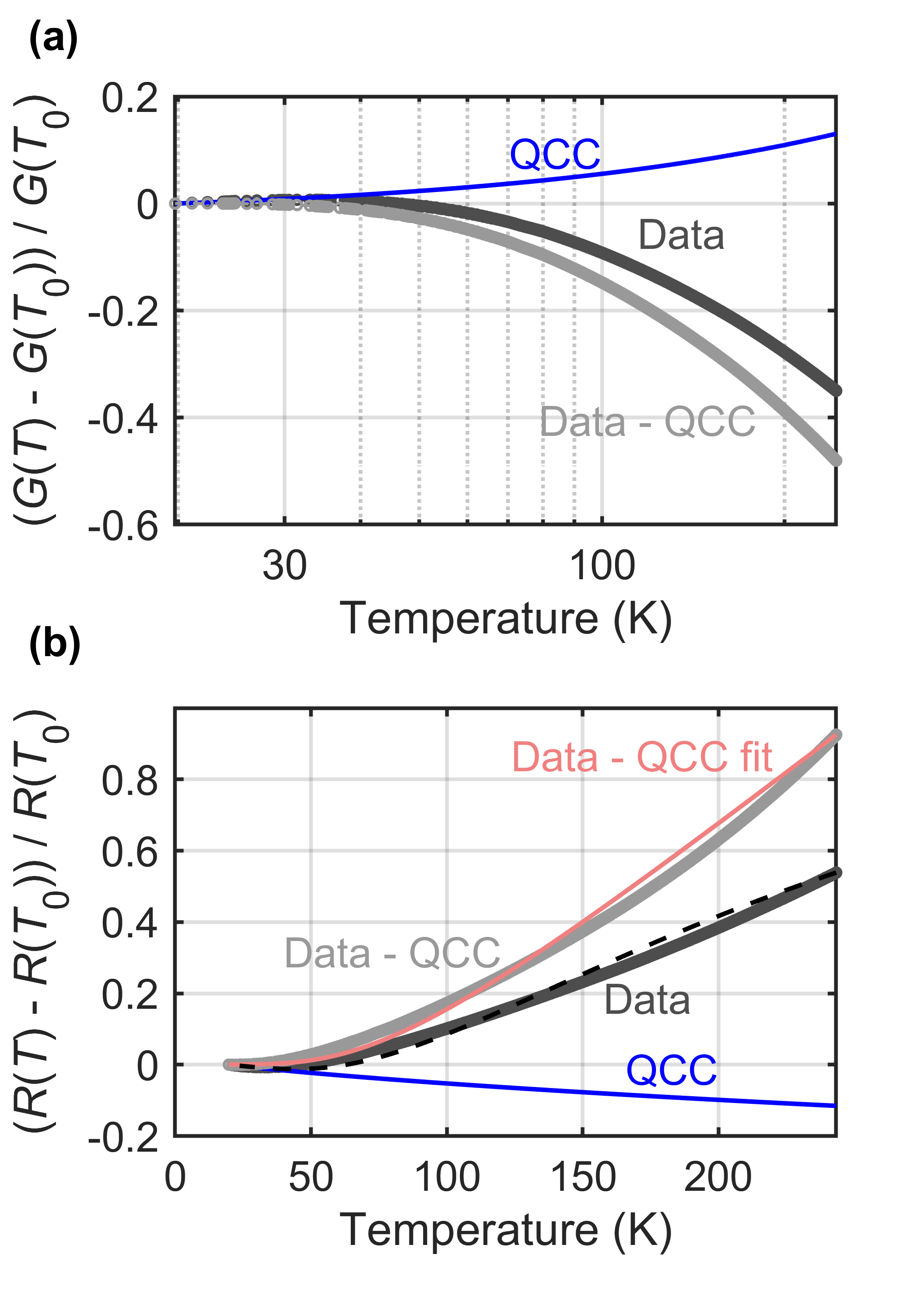}
\caption{\label{fig:wide}(a)~Normalized conductance as a function of temperature. Dark grey circular markers represent the experimental data; the blue solid line shows the QCC contribution; and the light grey markers show the residuals after subtracting the QCC model. (b)~Normalized resistance as a function of temperature. Light grey markers show the residuals; the light pink solid line is a fit using  Eq.\,\eqref{eq8}; the blue solid line is the QCC contribution; and the dark grey markers represent the original data.}
\label{Fig16}
\end{figure}

To analyze the high-temperature data, we applied the model proposed by Pop \textit{et al}.~\cite{Pop2007}, with the fitting process illustrated in Fig.\,\ref{Fig16}. Initially, we derived the quantum conductance correction (QCC) by fitting the low-temperature conductivity data using the 3D WL power-law temperature dependence, 
\[
\Delta\sigma_{\text{3D}}(T) \propto -T^{-3/2},
\]
which is consistent with established WL theory~\cite{Altshuler1985,lin2002}. This approach is reasonable because other mechanisms that contribute to the zero-field conductance, such as the Altshuler-Aronov (AA) corrections, exhibit power laws with temperature that are similar to those of WL. After deriving the combined QCC, as shown by the blue solid line in Fig.\,\ref{Fig16}(a), we removed this correction from the dataset, as shown by the small gray circles that overlap closely in Fig.\,\ref{Fig16}(a).

Subsequently, we transformed the $y$-axis to resistivity and utilized the equation for the resistance of individual metallic SWCNTs as
\begin{equation}
\begin{split}
      R_\text{m}=\frac{h}{4e^2}\left(1+\frac{L_\text{t}}{\ell_\text{eff}}\right)
\end{split}
\label{eq8}
\end{equation}
where $L_\text{t}$ is the length of an individual nanotube, taken as $L_t = 6\,\upmu$m based on the measured aspect ratio and the average tube diameter, and $\ell_\text{eff}$ is the effective carriers' mean free path (MFP), written as $1/\ell_\text{eff} = 1/\lambda_\text{AC} + 1/\lambda_\text{OP}^\text{ems} + 1/\lambda_\text{OP}^\text{abs}$, where $\lambda_\text{AC}$ is the elastic electron scattering by acoustic phonons, $\lambda_\text{OP}^\text{abs}$ is the inelastic optical phonon absorption, and $\lambda_\text{OP}^\text{ems}$ is the inelastic optical phonon emission. They are given as 
\begin{align}
     \lambda_\text{AC}(T) &= \lambda_\text{AC300}\frac{300}{T},\\
     \lambda_\text{OP}^\text{abs}(T) &= \lambda_\text{OP300}\frac{N_\text{OP}(300)+1}{N_\text{OP}(T)},\\
     \lambda_\text{OP}^\text{ems}(T) &= \lambda_\text{OP}^\text{abs}+\lambda_\text{OP300}\frac{N_\text{OP}(300)+1}{N_\text{OP}(T)+1},
\end{align}
respectively, where 
$\lambda_\text{OP300}\approx15$\,nm, and $N_\text{OP}$ is the optical phonon occupation written by Bose-Einstein statistics as
\begin{align}
    N_\text{OP}(T)=\frac{1}{\text{exp}\left(\frac{\hbar\omega_\text{OP}}{k_\text{B}T}\right)-1}.
\end{align}
The  light pink solid line in Fig.\,\ref{Fig16}(b) shows the fit of the residual data post-WL adjustment. This procedure yielded three fitting parameters: the optical phonon energy $\hbar \omega_\text{OP}$ at 0.023\,eV, acoustic phonon MFP $\lambda_\text{AC300}$ at $1.3\,\upmu$m, and optical phonon MFP $\lambda_\text{OP300}$ at $0.28\,\upmu$m, consistent with findings from a prior study~\cite{dini2020}. The comprehensive fit, represented by the black dashed line in Fig.\,\ref{Fig16}(b), demonstrates remarkable congruence with the original temperature dependency data, as shown by the dark grey markers in Fig.\,\ref{Fig16}(b).

\section{Discussion}
Our magnetotransport analysis reveals that electrical conduction in ultrahigh-conductivity CNT fibers arises from a complex interplay between 1D coherence along individual CNT bundles and diffusive coupling across a 3D network. The successful implementation of a hybrid 1D+3D WL model captures the full magnetic field and temperature dependence of the conductance, and aligns with the hierarchical structure observed in microscopy. The temperature-dependent crossover parameter $M(T)$ further confirms that inter-bundle coupling strengthens with increasing temperature, effectively driving a dimensional transition from quasi-1D to more isotropic 3D transport behavior.

A key finding is the observation of long phase coherence lengths, particularly $L_{\varphi,\text{3D}} \sim 180$\,nm at 1\,K, indicating an intrinsically low degree of disorder within the fibers. These coherence lengths are substantially larger than those reported in previous studies on CNT films and networks~\cite{Piraux2015,Wang2018}, highlighting the quality of the solution-spun, highly aligned, and ICl-doped CNT assemblies used here. Our approach to extract both 1D and 3D coherence lengths enables a more complete characterization of quantum interference effects in structurally anisotropic conductors.

Despite the good agreement between our model and experimental data, the necessity of a phenomenological scaling factor ($\sim$200) to match the magnitude of the quantum correction points to intrinsic limitations of the standard WL framework in describing mesoscale anisotropic systems. This discrepancy likely stems from inhomogeneous current distribution, nonuniform bundle packing, and anisotropic scattering rates that are not fully accounted for in conventional WL theory. Future theoretical work incorporating anisotropic diffusion tensors or percolative inter-bundle coupling could help bridge this gap.

Beyond weak localization, the high-temperature transport data were shown to follow a modified metallic transport model~\cite{Pop2007}, accounting for acoustic and optical phonon scattering. This dual-regime analysis—covering both quantum corrections and phonon-limited conduction—provides a comprehensive framework for understanding charge transport in macroscopic CNT materials.

Looking forward, our findings suggest that further improvement in alignment, doping uniformity, and interface engineering could push the coherence length even higher, opening the door to mesoscopic interference phenomena such as Aharonov--Bohm oscillations ~\cite{Bachtold1999,Cao2004} in fiber ring geometries. Moreover, the ability to tune dimensionality and coherence in scalable CNT fibers holds promise for next-generation quantum-coherent or low-loss energy transmission technologies based on flexible carbon conductors. A complete theoretical description of WL in these fibers will require accounting for anisotropic diffusion across the hierarchical structure, as well as the role of inelastic scattering in mediating decoherence between bundles.

\vspace{16pt}
\begin{acknowledgments}
We acknowledge the Shared Equipment Authority at Rice University for providing access to instrumentation used in this work, and thank the SEA staff for their assistance with sample preparation and materials characterization. J.K.\ acknowledges support from the Robert A.\ Welch Foundation through Grant No.\ C-1509 and the Air Force Office of Scientific Research through Grant No.\ FA9550-22-1-0382.
D.N.\ and L.C.\ acknowledge support from DOE BES DE-FG02-06ER46337 for magnetoresistance measurements.
L.W.T. was supported by the Department of Defense through a National Defense Science and Engineering Graduate (NDSEG) Fellowship, 32 CFR 168a.
G.W. and Y.S. acknowledge support from the Carbon Hub and from the National Science Foundation through Grant 2230727.
\end{acknowledgments}

\appendix
\vspace{16pt}
\section{Optical Characterization of CNT Films with Different Treatments}
\label{appendix:optical}

We performed optical spectroscopy on aligned CNT thin films representing three distinct post-spinning conditions: as-produced, ICl-doped, and annealed. These films were fabricated from the same CSA-based CNT solution used for fiber spinning, using the blade-coating method described in Ref.~\cite{HeadricketAl18AM}, followed by coagulation in acetone to preserve alignment.

The ICl doping was carried out via vapor-phase exposure under vacuum at $160^\circ$C, while annealing was conducted at $500^\circ$C in an inert atmosphere. All processing steps for the films mirrored those applied to the CNT fibers, ensuring consistency in comparing their electronic states.

\begin{figure}
\includegraphics[width=0.5\textwidth]{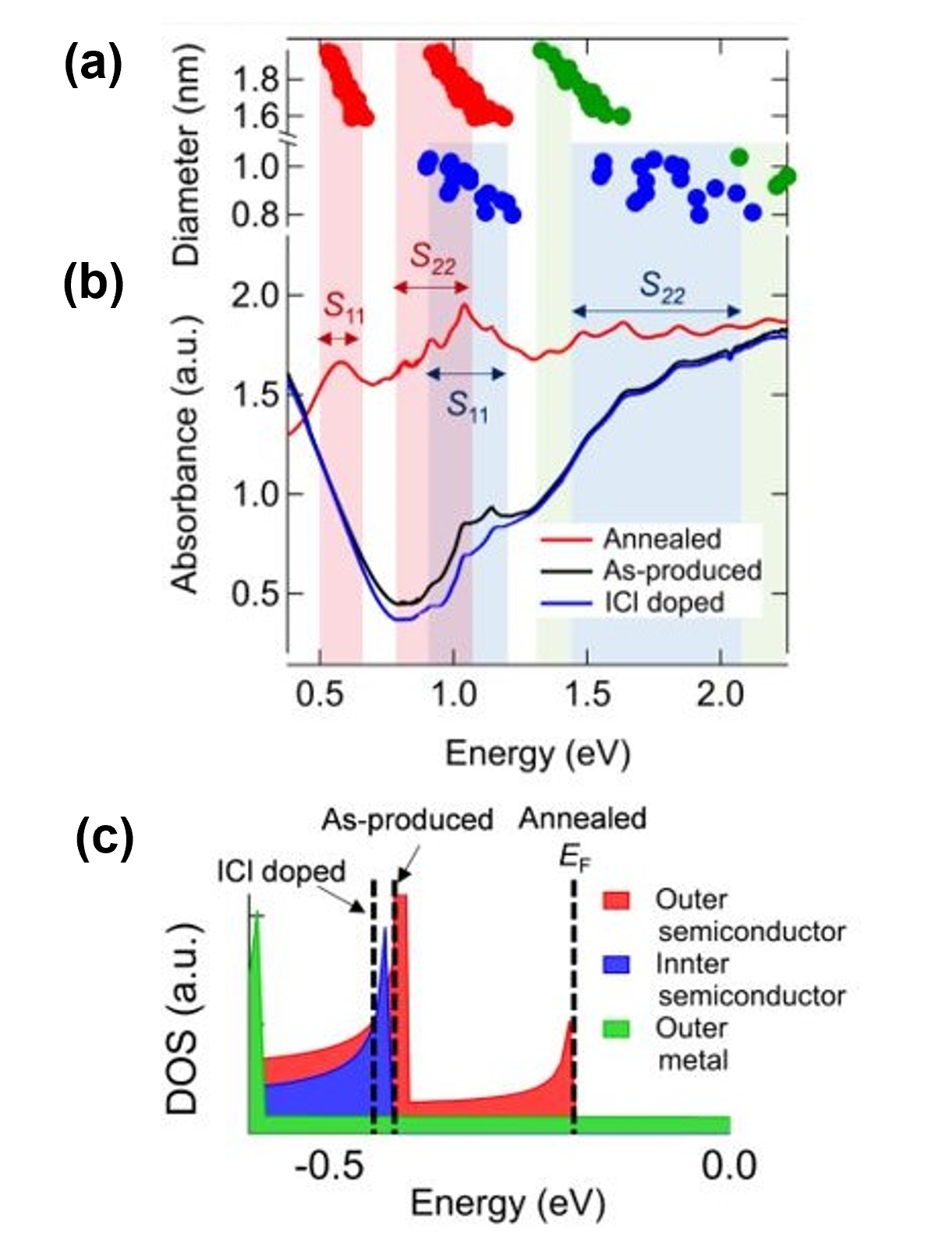}
\caption{(a) Expected peak positions based on diameter information, including transition peaks due to the outer semiconducting tubes (marked as $S_{11}$ and $S_{22}$, red), inner semiconducting tubes (marked as $S_{11}$ and $S_{22}$, blue), and outer and inner metallic tubes (green). 
(b) Absorbance spectra of three types of CNT films: as-produced (black), ICl doped (blue), and annealed at $500^{\circ}$C (red). 
(c) Schematic density of states (DOS) as a function of energy, for an outer semiconducting CNT (red), an inner semiconducting CNT (blue), and an outer metallic CNT (green). The expected $E_F$ position deduced from analysis of (b) for each sample is shown by the black dashed line.}
\label{fig:S1}
\end{figure}

Figure~\ref{fig:S1} summarizes the spectroscopic results. Panel (a) shows the expected optical transition energies based on the diameter distribution of CNTs, including $S_{11}$ and $S_{22}$ transitions from semiconducting species and the contributions from metallic tubes. Panel (b) presents the measured absorption spectra for the three treatment types. Compared to the as-produced state (black), ICl doping (blue) leads to suppression of the $S_{11}$ and $S_{22}$ transitions, indicating Fermi level shifting into the valence band. Annealing (red) causes partial recovery of optical transitions, suggesting dopant removal and partial bandgap reopening. Panel (c) provides schematic density of states (DOS) models corresponding to each treatment, with the inferred Fermi levels indicated by black dashed lines.

These results highlight the tunability of the CNT electronic structure through post-spinning treatment. While the main text focuses on ICl-doped fibers, Fig.~\ref{fig:S1} offers comparative insight into how chemical doping and thermal annealing modulate the density of states and Fermi level position.

\section{Estimation of $E_\textrm{F}$ from Optical Spectroscopy}

Figure~\ref{fig:S1}(b) compares the absorbance spectra of as-produced and ICl-doped CNT films, revealing distinct changes associated with Fermi-level shifts. In both samples, the suppression of the outer-tube $S_{11}$ and $S_{22}$ transitions can be attributed to Pauli blocking—that is, the inhibition of optical transitions when the final electronic states are already occupied due to the Pauli exclusion principle. The persistence of inner-tube $S_{11}$ peaks further suggests that the corresponding subbands remain partially unfilled. 

Based on the known correlation between nanotube diameter and excitonic transition energies~\cite{WeismanBachilo03NL}, the observed spectral evolution allows estimation of the Fermi energy positions. 
Figure~\ref{fig:S1}(c) schematically shows the estimated $E_\textrm{F}$ positions derived from these spectral data, indicating that for as-produced samples, $E_\textrm{F}$ lies near the first van Hove singularity (VHS) of the outer semiconducting CNTs, while for ICl-doped films, it shifts to the vicinity of the first VHS of the inner semiconducting CNTs, illustrating how post-spinning treatment modifies the electronic structure.

\bibliography{Natsumi}
\end{document}